\documentclass [12pt]{article}
\usepackage{epsfig}
\begin{document}
\title {Microcanonical Simulation of Complex Actions: The Wess Zumino Witten Case}
\author{Belal E. Baaquie \thanks{corbeb@nus.edu.sg} \\Y S
Seng \thanks{syeowsin@dso.gov.sg}\\Department of Physics\\National University
of Singapore\\Lower Kent Ridge Road\\Singapore 119260}
 \maketitle
 \begin{abstract}
 We present the main results of our microcanonical simulation of the Wess Zumino Witten
 action functional. This action, being highly non-trivial and capable of exhibiting
 many different phase transitions, is chosen to be representative of general
 complex actions. We verify the applicability of microcanonical simulation by
 successfully obtaining
 two of the many critical points of the Wess Zumino Witten action. The
 microcanonical
 algorithm has the additional advantage of exhibiting critical behaviour for
 a  small $8\times 8$ lattice. We also briefly discuss the subtleties
 that, in general, arise in simulating a complex action. Our
 algorithm for complex actions can be extended to the study of
 D-branes in the Wess Zumino Witten action.
 \end{abstract}
 \setcounter{equation}{0}
 \section {Description and Results}

In Minkowski spacetime,
the Feynman path integral has a complex integrand given by $\displaystyle e^{iS_M}$, where
$S_M$ is the Minkowski action. On Euclidean continuation of spacetime, the integrand
becomes real, namely, $\displaystyle e^{-S_E}$. The positive definite
Euclidean action $S_E$ is ideally suited for Monte Carlo simulations based on
the canonical ensemble. There is, however, a
class of actions that are generically of the form
$S=S_R+iS_I$ in {\it both} Minkowski and Euclidean spacetime. These actions are of
fundamental importance in physics, and we refer to
them as complex actions. For example,
all topological effects appear in the action in the form of
complex actions. A chemical potential term renders the action
complex. Lattice QCD at finite temperatures and and densities has a complex action,
and so do chiral gauge theories. Complex actions appear in study of non-Hermitian
hamiltonians, complex Langevin processes, random matrix models and
so on.

A complex valued action does not have any
probabilistic interpretation, and hence, for example the Metropolis algorithm
is unsuitable. All attempts to numerically
simulate complex actions for quantum field theories
have met with formidable difficulties \cite{wl,kg,ls} due to
diverging errors arising from increasingly large volumes
needed for obtaining the continuum limit. In particular, the conventional
Metropolis algorithm fails for complex actions due to
large errors in simulating the expectation values of cosines and sines
of extensive quantities. These expectation values
are necessary due to the presence of the imaginary component in the action.

There have been some studies that simulate complex actions using
the Langevin equation; however, the results of these simulations
cannot be shown to converge to the
required distribution \cite{ask,gl}. Other methods \cite{ag} also suffer from
either a restriction
to a very small  lattice, or are not efficient or well defined.

In our work, we work directly with the Feynman path integral
formulation of quantum field theory. We apply a generalized form
of the microcanonical method to the case of complex actions.
The microcanonical method has been very successful in simulating
global quantities of real statistical systems \cite{GB,MK,BEB}.
The microcanonical method yields a direct computation of the
partition function (up to a proportionality constant) and hence
the determination of its (complex) zeros. The zeros themselves contain useful
information, in particular whether
the system has critical points. The microcaconical simulation technique therefore
offers an attractive alternative to the usual methods, based on
the canonical ensemble,
for determining the phase diagram of statistical systems.

In this paper we extend the microcanonical algorithm to complex
actions, and, compute
its partition function. In particular, we study the well known and highly nontrivial
Wess Zumino Witten (WZW)
complex valued action functional. Since this model exhibits
many distinct phase transitions, we are able to
provide a solid test of our algorithm. Our simulations  {\bf predicts} the critical
parameters, and we confirm the effectiveness and accuracy
of our simulation by comparing the simulated results with the known
analytical results.

The Wess Zumino Witten  action in two-dimensional
Euclidean space has the following form
\begin{equation}
S=S_\lambda+ikS_I,
\end{equation}
where the chiral term is given by
\begin{equation}S_\lambda=\frac{1}{4\lambda^2}\int
d^2x
Tr(\partial_\mu g\partial_\mu g^{-1})\end{equation} \\
and the imaginary piece comes from the Wess Zumino  term
\begin{equation}
S_I=\frac{1}{24\pi}\int\limits_{S_3}d^3y\epsilon_{ijk}Tr[\bar{g}^{-1}\frac{\partial{\bar{g}}}
{\partial{y^i}}\bar{g}^{-1}\frac{\partial\bar{g}}{\partial
y^j}\bar{g}^{-1}\frac{\partial\bar{g}}{\partial
y^k}].
\end{equation}
For the quantum field theory to be well defined, $k$ must be integer-valued.

For the case of $SU(2)$, the Wess Zumino  term
can be parametrized in terms of the Euler angles \cite{Witt},
namely,
\begin{equation}S_I=\frac{1}{\pi}\int
d^2x\phi(x)sin^2\psi(x)sin\theta(x)\epsilon^{\mu\nu}\partial_\mu\psi(x)
\partial_\nu\theta(x).\end{equation}

As is well known, the WZW model has infinitely many critical points, one for each integer
value for $k$. The WZW model is conformally invariant
at the critial coupling constants given by\cite{Witt}
\begin{equation}\lambda_c^2=\frac{4\pi}{k}~~~;~~\mathrm{k:integer}.\end{equation}

By extending the Lee-Yang
theorem for zero's of the partition function to the complex case, we expect that the
partition function, in the infinite volume limit, will be zero close to the critical
coupling constants. We should be able to locate a zero (close to the continuum model) if
the size of the lattice is large enough.

By calculating the simulated
partition function as a function of $k$, the zeroes can be identified
as the points of the $k$ axis where the partition function $Z$ crosses over from a
positive to a negative value. This cross over is a
unique feature of complex actions, since the partition
function of a real action can never be negative. Given that only
integer $k$'s are physical, the only physical zeros of $Z$ are
those that are close to integer values of $k$.

The partition function of the Wess Zumino Witten action functional
is given by
\begin{equation}
Z(\lambda,k)=\int[DU]\exp(-S_\lambda)\exp(ikS_I),
\end{equation}
which in turn can be cast into the form
\begin{equation}Z(\lambda,k)=\int_{-\infty}^{+\infty}{dE \rho(\lambda,E)\exp(ikE)}
\end{equation}
with the density of states given by
\begin{equation}\rho(\lambda,E)=\int{[DU]\delta(E-S_I)\exp(-S_\lambda)}
\end{equation}

The density of states is
estimated by a generalization of the microcanonical algorithm. First, the
simulation space is broken up into energy
bins of size $\triangle{E}$. Bin $E=0$ is taken to be the
reference bin. The system is thermalized by randomly generating
configurations and accepting only those configurations whose
values of $S_I$ fall within the first two bins. The density of
states $\rho(\lambda,\triangle{E})$ is estimated by calculating
the occupation number ratio between the two bins. The updating of
configurations and crediting of counts to the bins are based on
a {\it combination} of the microcanonical and canonical ensembles. This
is done as follows.

\begin{figure}[h]
\begin{center}
\epsfig{file=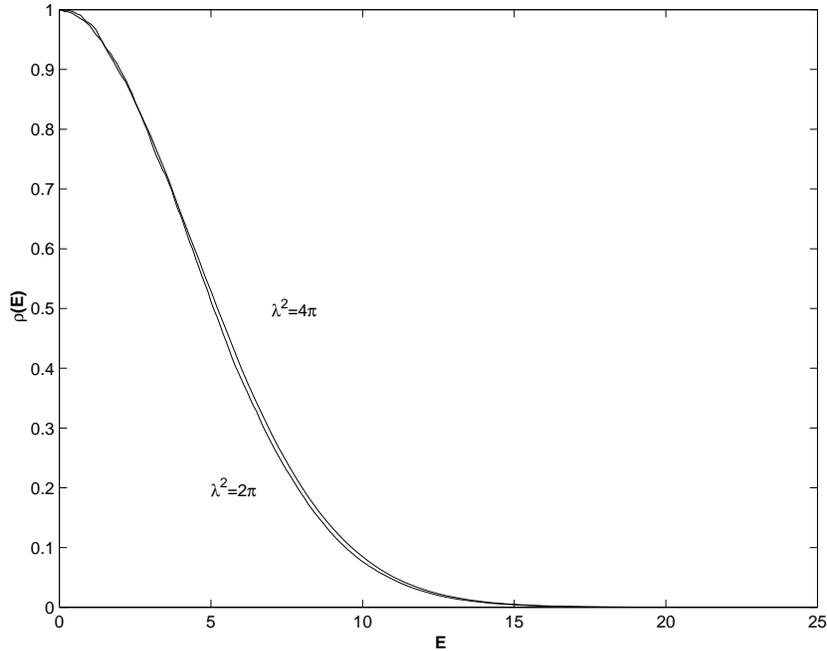, width=11cm}
\caption{Density of states for two values $\lambda_c$}
\end{center}
\label{one}
\end{figure}

\begin{enumerate} \item A new configuration is generated, and its
acceptance is weighted by the usual Metropolis factor
$\exp(-S_\lambda)$. Note that since the chiral contribution, being real, plays
the role of a weighting action, the value of $\lambda$ has to be at the {\bf critical
value} for a particular level; otherwise, if $\lambda \neq \lambda_c$, the system
would never go critical.
\item The updated configuration is accepted
only if it's $S_I$ value falls within the two bins.
If this is so, the counter for the bin where $S_I$ lands
is credited. \item If the updated configuration does not satisfy
the above criterion, a count will still be credited to the bin of the
{\bf old} configuration.\end{enumerate}

\begin{figure}[h]
\begin{center}
\epsfig{file=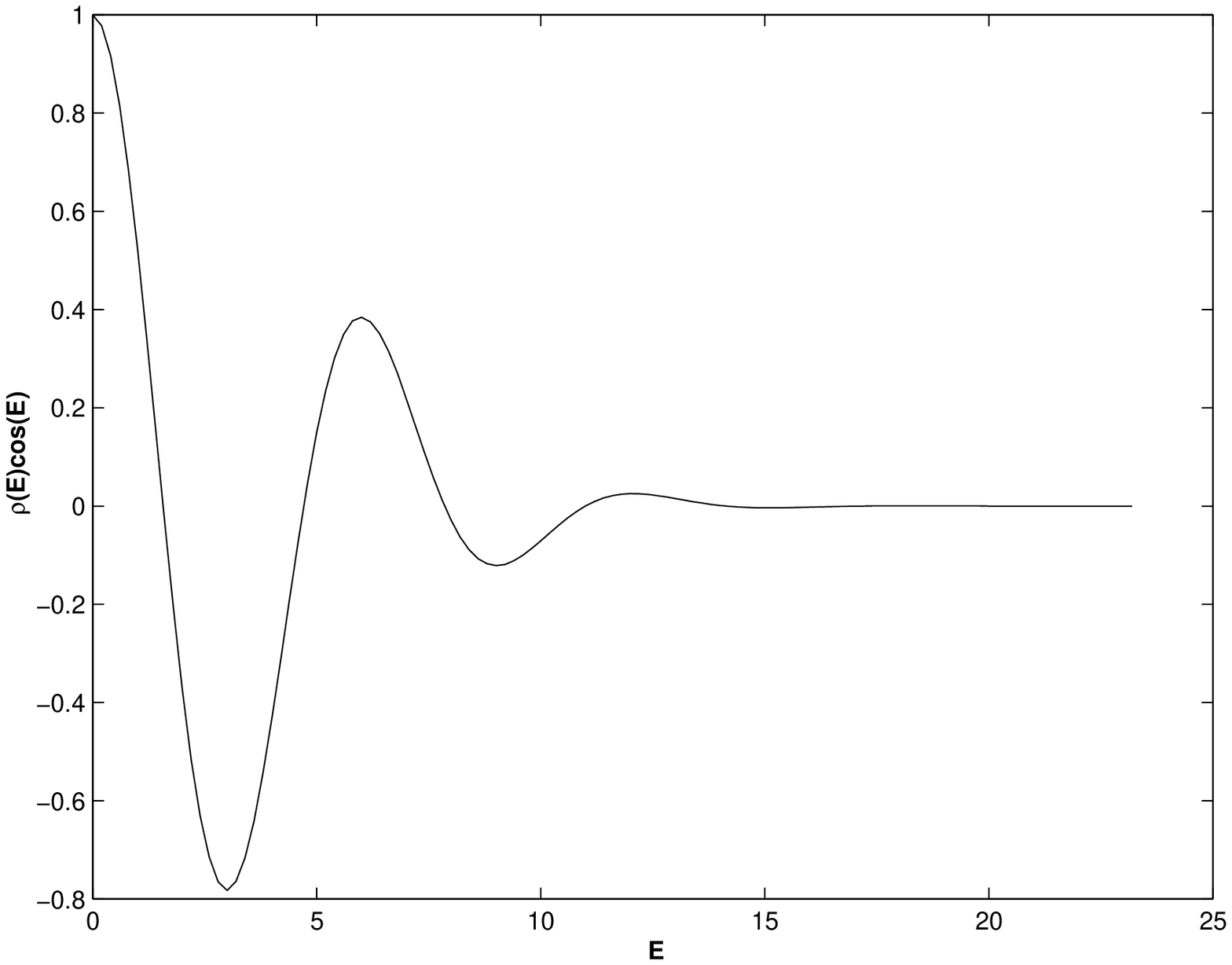, width=11cm}
\caption{The integrand of $Z$ for $\lambda_c=4\pi$}
\label{twoa}
\end{center}
\end{figure}

\begin{figure}[h]
\begin{center}
\epsfig{file=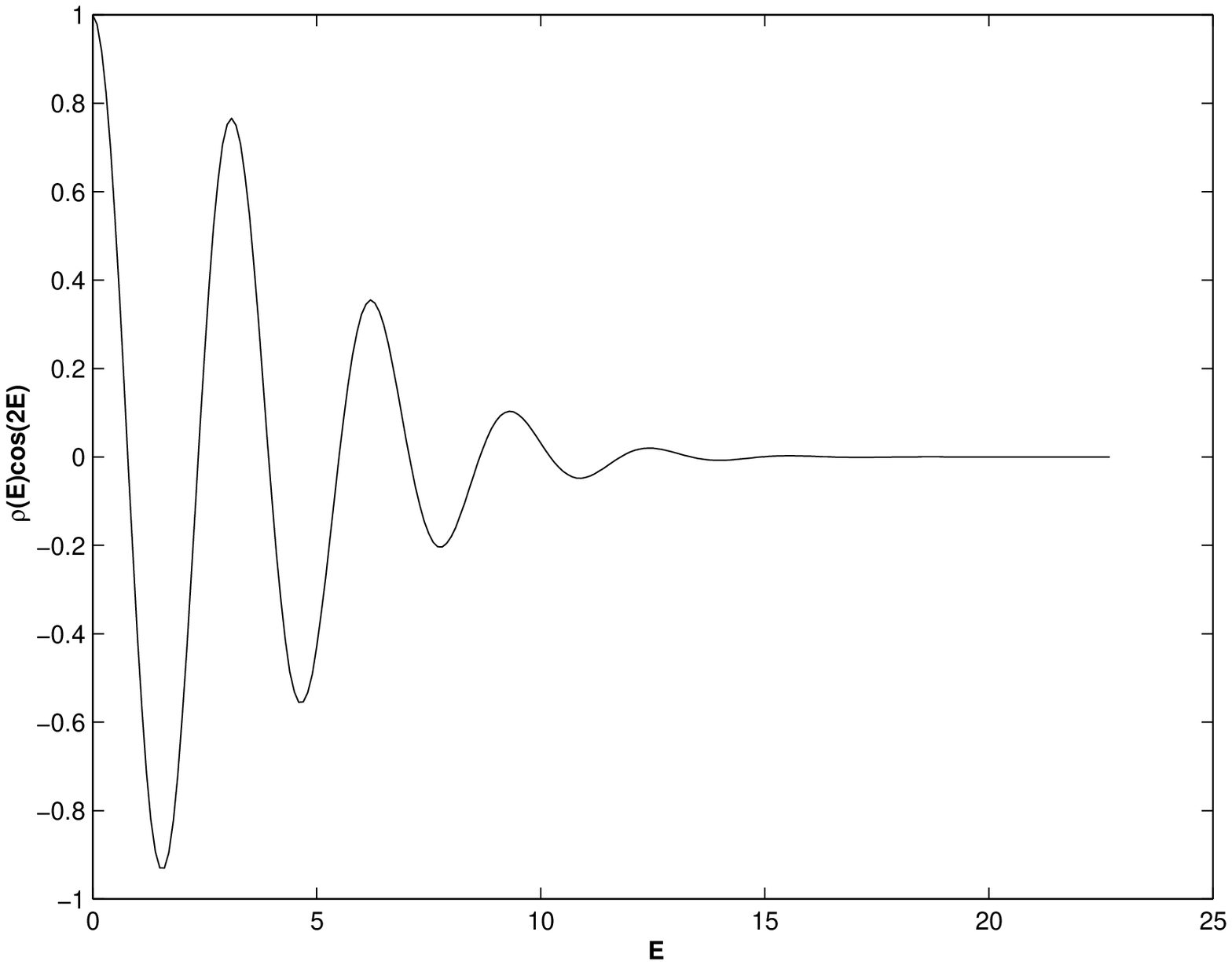, width=11cm}
\caption{The integrand of $Z$ for $\lambda_c=2\pi$}
\label{twob}
\end{center}
\end{figure}

\begin{figure}[h]
\begin{center}
\epsfig{file=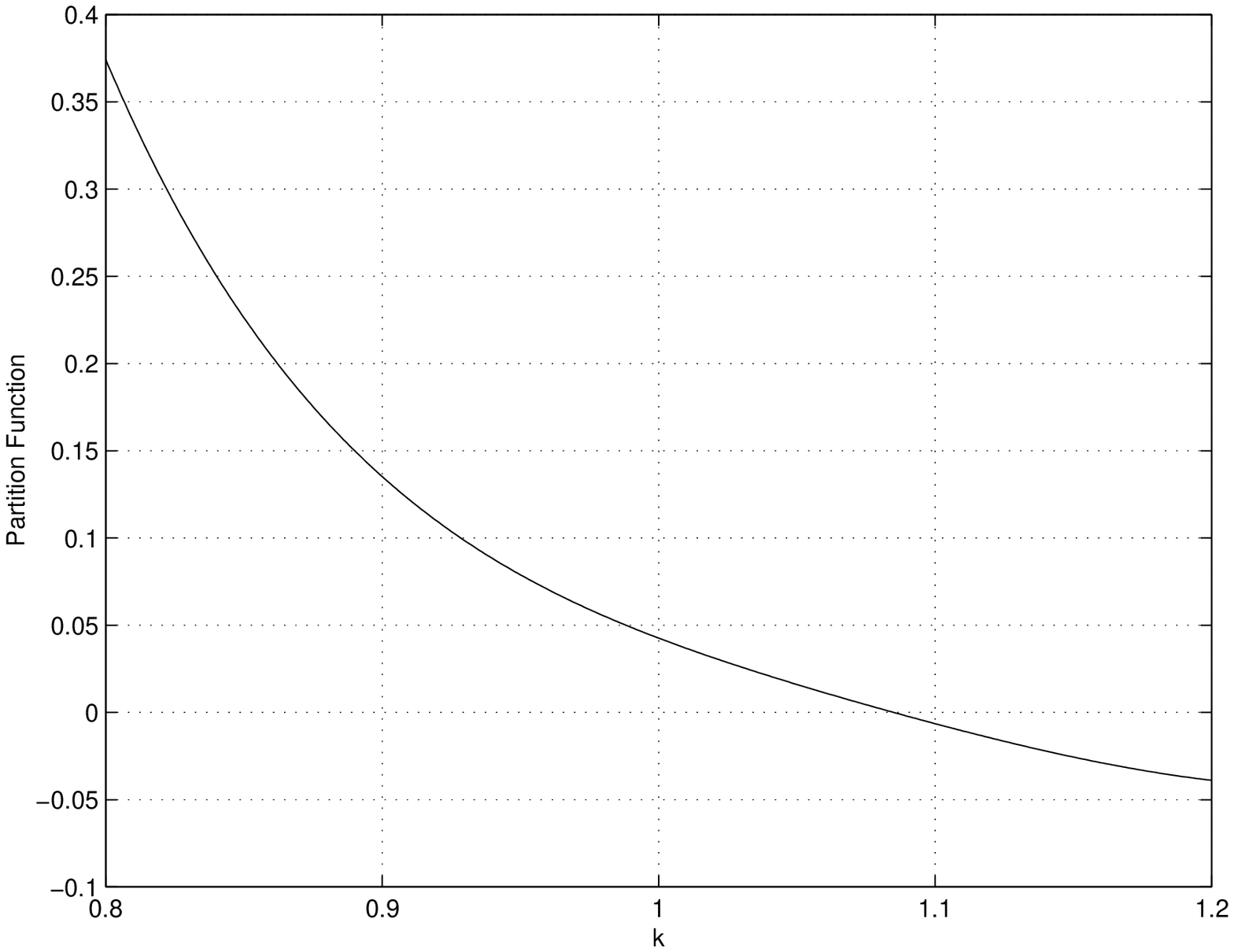, width=11cm}
\caption{Scan of $Z$ near a zero for $\lambda_c=4\pi$. The zero occurs at $k\simeq1.08$.}
\label{threea}
\end{center}
\end{figure}

\begin{figure}[h]
\begin{center}
\epsfig{file=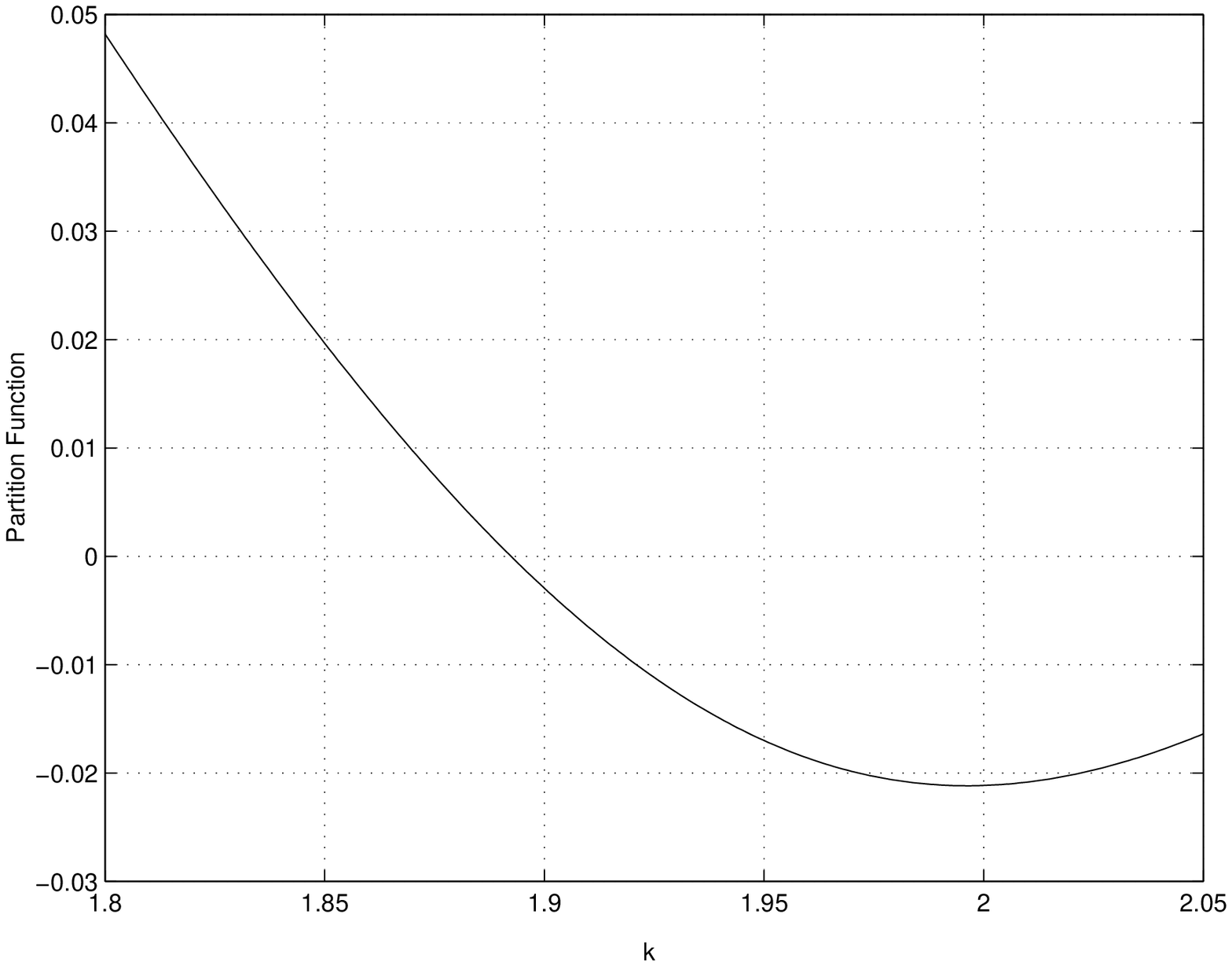, width=11cm}
\caption{Scan of $Z$ near a zero for $\lambda_c=2\pi$. The zero occurs at $k\simeq1.89$.}
\label{threeb}
\end{center}
\end{figure}

By continuing this process across all the energy bins allows us to get
an estimation of the density of states. A crucial point in
going from one set of bins to another is to have, in between, at least one overlapping
bin for the purpose of normalizing the occupation
number of the various bins.

All of the results quoted are for simulations for a $8\times 8$
lattice, as this size is sufficient for accurately estimating the value of
critical $k$. In our work, we tested our simulations for two levels, namely $k=1$ and
$k=2$.  For the critical coupling $\lambda^2_c=4\pi, \triangle E$ was set
to be 0.2 while a finer value of 0.1 was chosen for
$\lambda^2_c=2\pi$. A total number of about 1 million
configurations were generated for every set of energy bins. The
energy bin was terminated when the resulting density of state is
less than 3 decimal places (which is the achieved level of
accuracy in our simulations).
The results for the simulated density of states is shown in Figure \ref{one}.

Since $\rho(\lambda,-E)=\rho(\lambda,E)$, it is
only necessary to perform the simulation for
$E\geq 0$. This reflection symmetry
ensures that the WZW partition function is real, since it simplifies to
\begin{equation}Z(\lambda,k)=2\int_{~0}^{\infty}{dE\rho(\lambda,E)cos(kE)}
\end{equation}
The simulated value for integrand of $Z(\lambda_c,k)$ is given in Figures \ref{twoa}
and \ref{twob}. The plot of $Z(\lambda_c,k)$ against $k$ is physical
relevant
only for integer $k$'s. We expect that $Z=0$ at the value of $k$ that corresponds to
$\displaystyle \lambda_c^2=\frac{4\pi}{k}$; for other values of
integer $k$, $Z$ is physical and contains  information on
how the theory behaves away from criticality.

The behaviour of $Z$ near its zero is given in Figures \ref{threea} and \ref{threeb}.
By scanning for zeros of the simulated partition function Z that are close to
integers, we find that:
\[k_{\mathrm{simulated}}=1.06\pm0.07 ~~~(\mathrm{for} ~\lambda^2=4\pi)\] and
\[k_{\mathrm{simulated}}=1.95\pm0.06 ~~~(\mathrm{for} ~\lambda^2=2\pi).\]
The errors were deduced from the standard deviation computed from two separate sets of
results for each coupling.

The exact values are $k=1$ and $k=2$ respectively, and we see that
we have obtained a result of fairly high accuracy with a surprisingly
small lattice.

\section {Discussion}
Our simulations of the Wess Zumino Witten action demonstrates that the
microcanonical method is in principle capable of simulating
complex actions. Unlike the case of real actions, however, there
are several subtleties involved. Chief among them are the
following.
\begin{enumerate}
\item A high degree of accuracy is required in order to achieve
stable and consistent results. The reason being that, unlike real
actions, terms in the complex action partition function  {\bf
cancel}
each other (see the graphs of Figures \ref{twoa} and \ref{twob}),
leaving behind a small value in the region where the phase transitions
take place.
\item The fact that the critical point can be made to move from $k=1$
to $k=2$ by changing the value of $\lambda_c$ in the weighting
function is strong evidence that our simulation is giving the
correct result. There are infinitely many critical points of the
WZW model, one for each integer $k$, and in principal, all these
critical points can be obtained by our simulation.
\item There is an additional necessity of having to provide a good
resolution of the $\cos(kE)$ term. Typically, this means that there should
be at least a few points within each trigonometrical quadrant.
Consequently, since $\displaystyle \triangle{E}\sim \frac{1}{k}$, a smaller
energy spacing is required for larger couplings. Computational
time therefore is expected to increase as coupling $k$ increases.
\item Since the cosine of an argument cannot be taken as a
reliable weight in any Monte Carlo simulation, the contribution
to the weight must always come from the real action. This constraint demands
that the coupling $\lambda$ in the real action has to be fixed at its critical
value $\lambda_c$. In other words, it
is not possible to fix the coupling in the imaginary component,
namely $k$, and deduce the critical coupling $\lambda_c$ of the real action.
\item Our method is ideally suited for studying the critical
properties of extended systems. In particular, D-branes in the WZW
model are non-local background configurations that respect the conformal
invariance of the theory \cite{AS}. Our method, for example, can numerically study
under what circumstances a D-brane background in the WZW model results in
a conformally invariant system.
\end{enumerate}

In conclusion, we believe that the ability of the microcanonical
technique to get accurate predictions from the Wess Zumino Witten action
is a major step forward in the simulation of complex actions, and opens the
way for numerically studying other complex actions, as well as extended objects
such as D-branes.
\section{Acknowledgements}
We have benefited from many discussions and communications with Gyan Bhanot.


\begin{thebibliography}{10}
\bibitem {wl}W V Linden, Phy. Rep. {\bf 220}, 53 (1992)
\bibitem {kg} T D Kieu, C J Griffin {\it Monte Carlo Simulations with Indefinite
and Complex-Valued Measure} hep-lat/9311072
\bibitem {ls} L I Salcedo, Phy. Lett. {\bf 304B}, 125 (1993)
\bibitem {ask} A S Kronfeld, {\it Dynamics of Langevin Simulation} hep-lat/9205008
\bibitem {gl} H Gausterer, S Lee {\it The Mechanism of Complex Langevin Simulation}
 hep-lat/9211050
\bibitem {ag}A Gocksh, Phys. Lett. {\bf 206B}, 290 (1991);
S B Fahy and D R Mamann, Phys. Rev. Lett. {\bf 65}, 3437 (1990).
\bibitem{GB} G Bhanot,K Bitar, P Carter,R Salvador,R Toral,
Physical Review Letters, {\bf 59}, 803, 1987.
\bibitem{MK} M Karliner, R Sharpe, Y K Chang, Nuclear Physics B,
{\bf 302}, 204,1988.
\bibitem{BEB} G Bhanot,B E Baaquie, Nuclear Physics B,{\bf 382},
409, 1992.
\bibitem{Witt} Edward Witten, Comm Math Phys., 92, 455-472, 1984.
\bibitem{AS} A Y Alexseev, V Schomerus {\it D-branes in WZW model} hep-th/9812193 v2
\end{thebibliography}
\end{document}